\documentclass{PoS}

\title{First Double Cascade Tau Neutrino Candidates in IceCube and a New Measurement of the Flavor Composition }

\ShortTitle{First Double Cascade Tau Neutrino Candidates in IceCube}  

\author{
The IceCube Collaboration\footnote{For collaboration list, see PoS(ICRC2019) 1177.}\\
{\itshape \href{http://icecube.wisc.edu/collaboration/authors/icrc19_icecube}{http://icecube.wisc.edu/collaboration/authors/icrc19\_icecube}}\\
E-mail: \email{juliana.stachurska@icecube.wisc.edu}
}

\abstract{The IceCube Neutrino Observatory at the South Pole, which detects Cherenkov light from charged particles produced in neutrino interactions, firmly established the existence of an astrophysical high-energy neutrino component. The expected neutrino flavor composition on Earth is $\nu_e:\nu_{\mu}:\nu_{\tau}$ of about 1:1:1 for neutrinos produced in astrophysical sources through pion decay. A measurement of the flavor composition on Earth can provide important constraints on sources and production mechanisms within the standard model, and can also constrain various beyond-standard-model processes. Here the measurement of the flavor composition performed on IceCube's High-Energy Starting Events sample with a livetime of about 7.5 years is presented. IceCube is directly sensitive to each neutrino flavor via the single cascade, track and double cascade event topologies. In IceCube, $\nu_{\tau}$-CC interactions above $\sim$ 100~TeV can produce resolvable double cascades, breaking the degeneracy between $\nu_e$ and $\nu_{\tau}$ present at lower energies. IceCube's first two identified double cascades are presented and the properties of the two $\nu_{\tau}$ candidates are discussed. \\

\vspace{4mm}
{\bfseries Corresponding author:}
 \speaker{J.Stachurska}\\
{\itshape DESY, Platanenallee 6, D-15738 Zeuthen, Germany}
}

\FullConference{36th International Cosmic Ray Conference -ICRC2019-\\
		July 24th - August 1st, 2019\\
		Madison, WI, U.S.A.}

\usepackage{textcomp}
\usepackage{gensymb}
\usepackage{cancel}
\usepackage{url}

\newcommand{\nutau}{\nu_{\tau}}
\newcommand{\numu}{\nu_{\mu}}
\newcommand{\nue}{\nu_{e}}
\newcommand{\nuratio}{\nu_{e}:\nu_{\mu}:\nu_{\tau}}

\newcommand{\diff}[1]{\mathrm{d}#1}
\DeclareMathAlphabet{\mathpzc}{OT1}{pzc}{m}{it}

\begin{document}

\section{Introduction}
The IceCube Neutrino Observatory \cite{IceCube} reported the discovery of a diffuse astrophysical neutrino flux in 2012 using the High-Energy Starting Events (HESE) selection \cite{IC1}. 
The HESE selection is an all-flavor event selection, with  events previously being classified by eye as cascade-like or track-like. In the newest iteration of HESE with 7.5 years of data, older IceCube data have been reprocessed and re-analyzed following a new detector calibration, and new data have been added. Further improvements have been made to the atmospheric background modeling and to the likelihood used for the analyses. Details as well as results of the diffuse flux measurement can be found in \cite{HESE7}. We have further incorporated an algorithmic topology identification \cite{Marcel} into the HESE processing, and performed a new flavor composition measurement. We classify two events as double cascades for the first time, yielding the first candidates for astrophysical tau neutrinos. The results were first presented in \cite{Neutrino}. After summarizing the results, we here present the details of follow-up work, aimed at assessing the probability of each of the two candidates to stem from a $\nutau$ charged current (CC) interaction in greater detail.

\section{The Flavor Composition in the High-Energy Starting Events}
There are three event topologies in IceCube to be distinguished and matched to the three neutrino flavors for a flavor composition measurement. Light depositions along an infinite track traversing the detector are called tracks and stem from $\nu_{\mu}$-CC interactions and atmospheric muons, as well as $\nu_{\tau}$-CC interactions where the $\tau$ lepton decays to a muon. Single cascades consist of one energy deposition and are produced by $\nu_{e}$-CC and all flavor neutral current (NC) interactions. Double cascades are two energy depositions connected in space and time by a (dim) track. They are produced by $\nutau$-CC interactions where the first cascade comes from the neutrino interaction and $\tau$ production, and the second from the $\tau$ decaying electromagnetically or hadronically. However, the short tau decay length of $\langle L_{\tau} \rangle \sim 50$~m $\cdot \, E_{\tau}$ / PeV, where $ E_{\tau}$ is the $\tau$ energy, makes the distinction between single and double cascades very challenging in IceCube, where the mean horizontal distance between sensors is 125~m. In IceCube, the double cascade topology opens up at deposited energies $\mathcal O$(100~TeV), breaking the degeneracy between $\nu_e$ and $\nu_{\tau}$ flavors present at lower energies.
All events are reconstructed using maximum likelihood fits for different hypotheses: single cascade \cite{Ereco}, track \cite{SPE} and double cascade \cite{Ereco, Patrick}. For the fits, the timing and spatial information of all light collected in an event is used. 
For a double cascade reconstruction to be deemed of sufficient quality, the fit must converge, a minimum of 1~TeV reconstructed energy is required for each of the cascades and the opening angle between the reconstructions using a double cascade and a track hypothesis cannot exceed 30\textdegree. Additionally, neither cascade can be more than 50~m outside of the instrumented volume. This soft containment requirement becomes increasingly constraining with increasing tau decay length, leading to a drop of successfully identifiable double cascades above lengths of $300-400$~m. 
Observables used for classification are the distance between the reconstructed cascades (called \textit{double cascade length} hereafter), the asymmetry between the two cascades' energies (called \textit{energy asymmetry} hereafter), and the fraction of the total energy deposited close to the cascade vertices (called \textit{energy confinement} hereafter). The energy asymmetry is defined as $A_E=(E_1-E_2)/(E_1+E_2)$, where $E_{1,2}$ are the reconstructed energies of the two cascades, and can take values $-1 \leq A_E \leq 1$. Events passing the quality, length, energy asymmetry and energy confinement cuts given in Table \ref{tab:cuts} are classified as double cascades. While the length resolution is $\sim$ 2~m, the minimum required length for double cascades is 10~m. The larger the minimum required length, the smaller is the contamination by true single cascades, but also the signal expectation drops rapidly above lengths of $\sim 20-30$~m. 
Events failing any of the double cascade cuts are classified as indicated in Table \ref{tab:sorting}.

\begin{figure}[tbh]
	\centering
	\includegraphics[width=100mm]{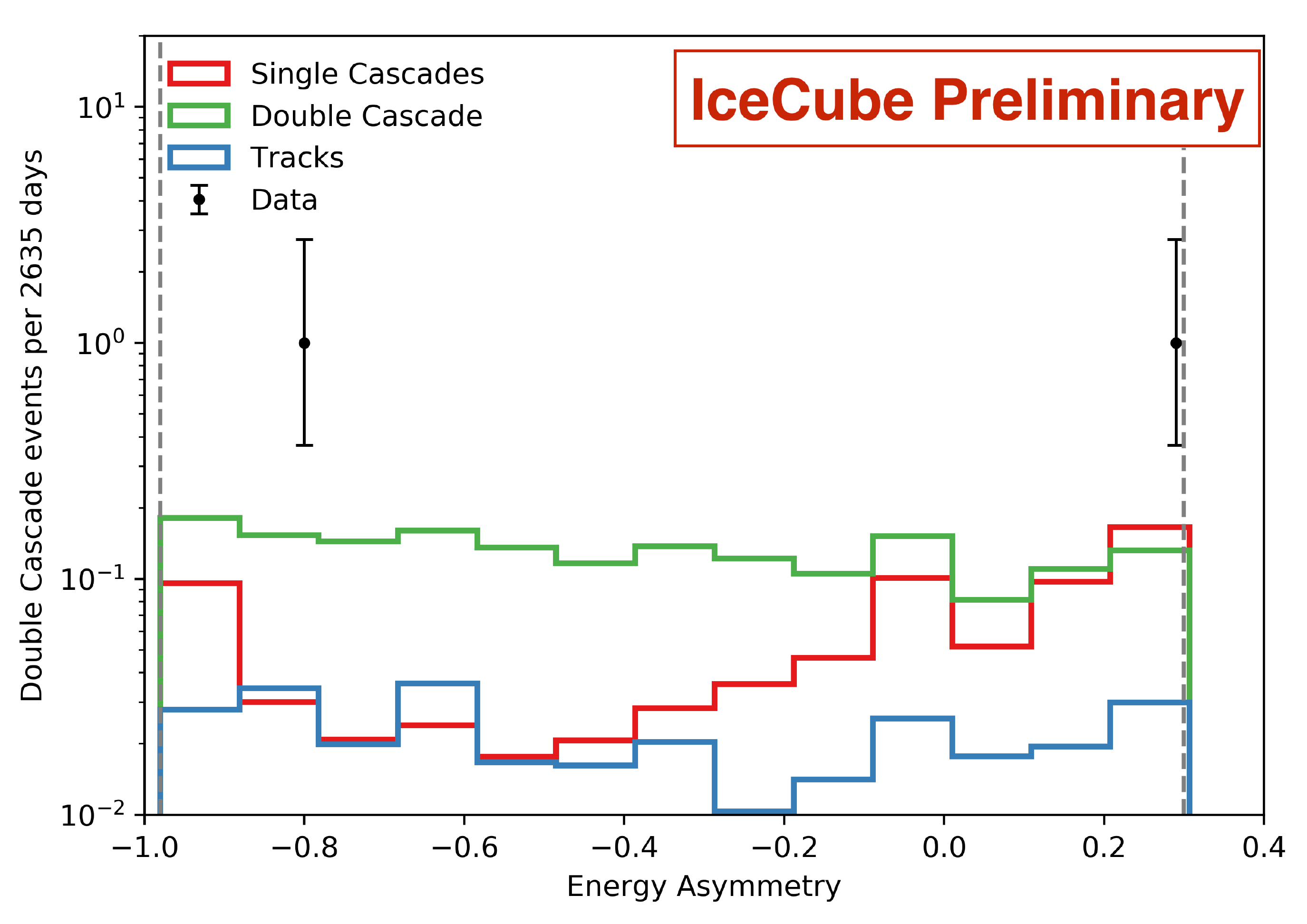}
	\caption{Distribution of the reconstructed energy asymmetry in the double cascade sample split by true topology for the best-fit astrophysical and atmospheric spectra based on simulation. The two double cascades (``Double Double'' at -0.80, ``Big Bird'' at 0.29) are shown.}
\label{fig:EA}
\end{figure}

\begin{table}[t]
\centering
\begin{tabular}{l | cc}
Observable  & Min & Max  \\\hline
Quality & \multicolumn{2}{c}{good} \\
Length & 10 m & --\\ 
Energy Asymmetry & -0.98 & 0.30 \\
Energy Confinement & 0.99 & 1.0 \\ \hline
\end{tabular}
\caption{Double cascade cuts. Note that while there is no upper limit on the length, the requirement that both cascades need to be contained within 50~m of the detector boundaries leads to an effective cutoff above $\sim 300-400$~m.}
\label{tab:cuts}     
\end{table}
\begin{table}[tbh]
\centering
\begin{tabular}{l | r}
Failed double cascade cut & resulting classification \\ \hline
Quality  &  depending on fit likelihoods \\
Length & single cascade \\
Energy Asymmetry & single cascade \\
Energy Confinement & track \\ \hline
\end{tabular}
\caption{Track and single cascade classification. The cuts are given in order of precedence, such that the first failed cut determines the resulting classification. For events failing the double cascade fit quality cuts, the likelihoods of the track and single cascade fits are compared, and the topology with the corresponding fit with the higher likelihood is chosen.}
\label{tab:sorting}     
\end{table}
\begin{figure}[b]
	\centering
	\includegraphics[width=130mm]{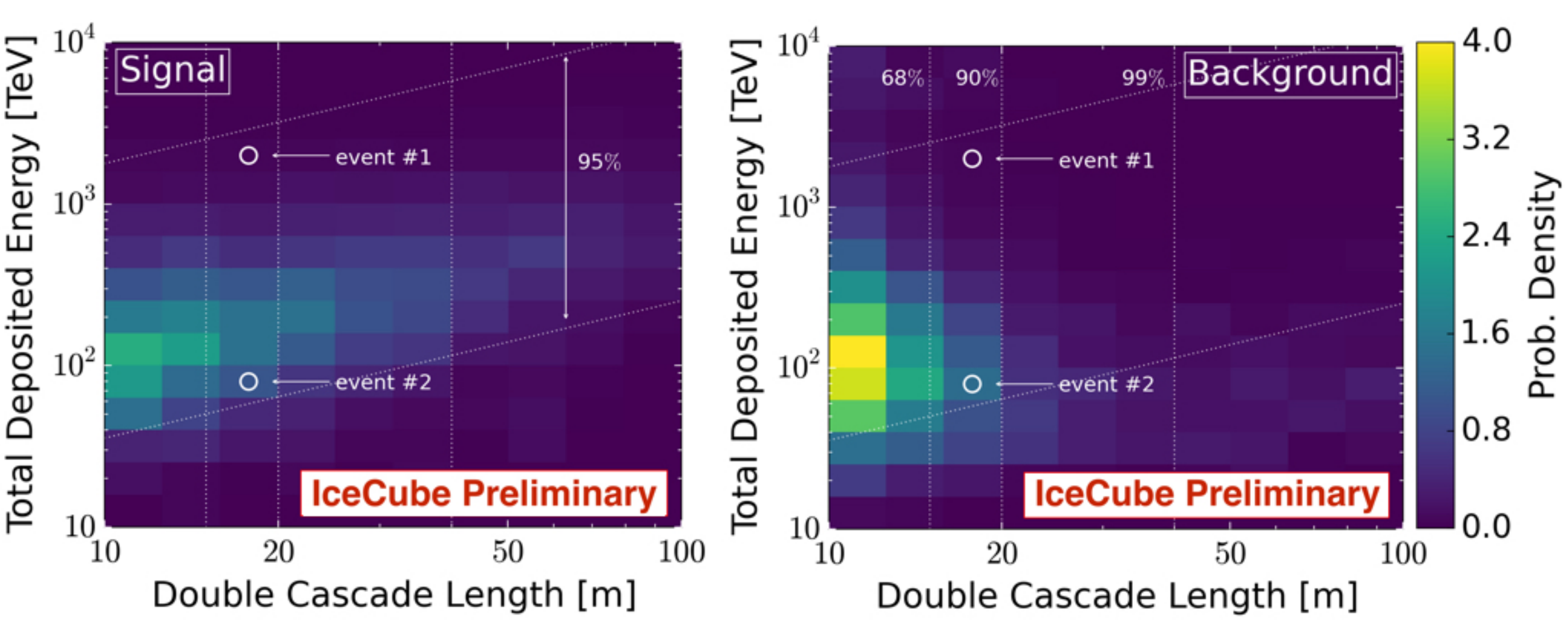}
	\caption{Total deposited energy against reconstructed double cascade length for the double cascade sample. Signal ($\nu_{\tau}$ -induced double cascade events) histogram (left). Background (all remaining events) histogram (right). The two tau-neutrino candidate events are overlaid as white circles.}
\label{fig:PID}
\end{figure}
\begin{figure}[tb]
	\centering
	\includegraphics[width=100mm]{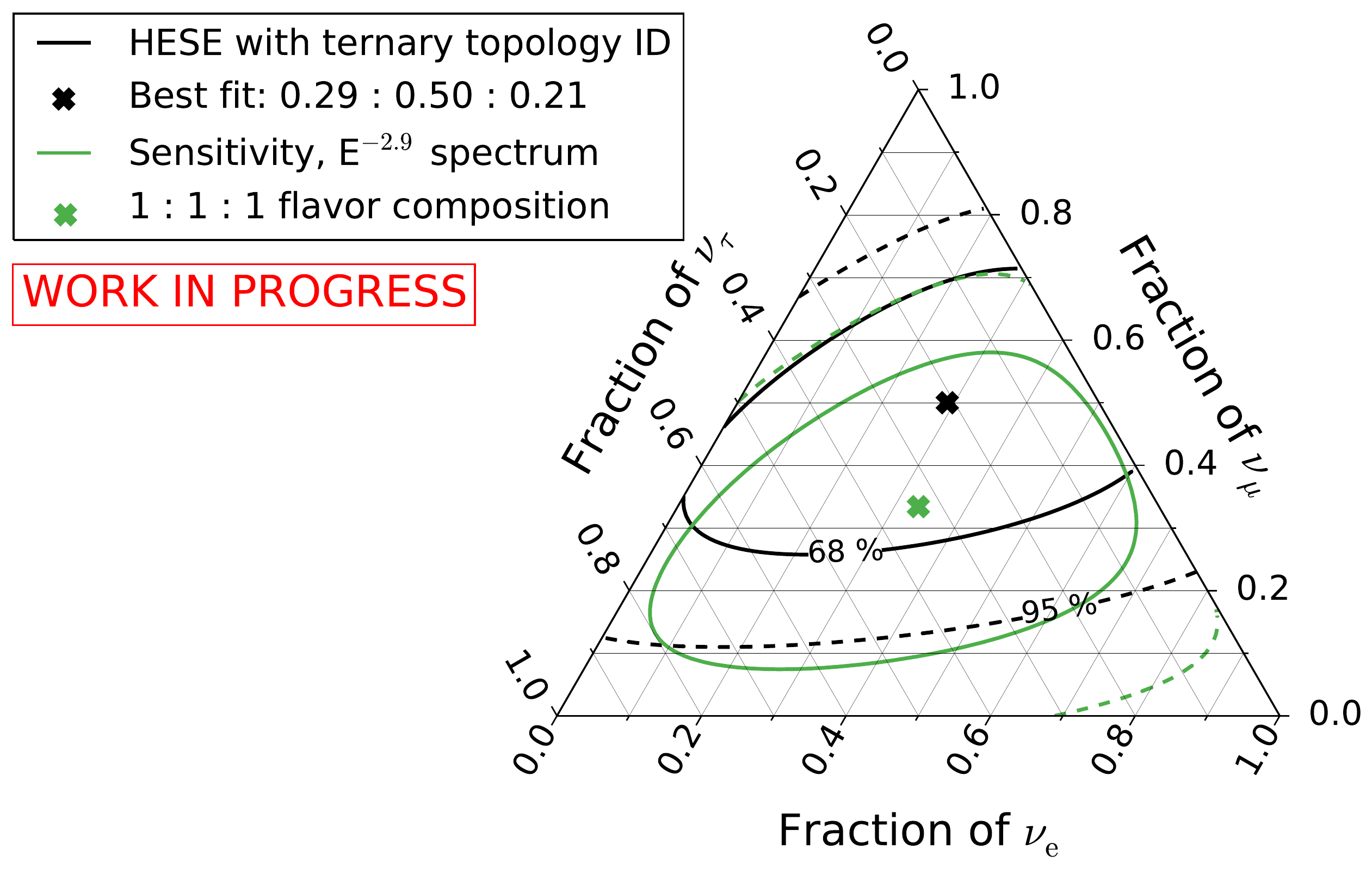}
	\caption{Measured flavor composition of IceCube HESE events with ternary topologyID and sensitivity at the best fit spectrum. Contours obtained using Wilks' theorem \cite{Wilks}.}
\label{fig:flavor3}
\end{figure}
Above 60~TeV reconstructed deposited energy, the HESE sample contains 60 events which we classify into 42 single cascades, 16 tracks, and 2 double cascades.
A multi-component maximum-likelihood fit is performed on the three topology samples using two-dimensional PDFs obtained from Monte Carlo simulations, as shown in Fig. \ref{fig:PID} for the double cascade topology sample. The likelihood used is the SAY-likelihood \cite{SAY}, developed for the 7.5 year HESE update, which takes limited Monte Carlo statistics into account.
For single cascades and tracks, the observables \textit{total deposited energy} and \textit{cosine $\theta$} are used, where $\theta$ is the zenith angle. For double cascades, total deposited energy and double cascade length are used. For $\nu_{\tau}$ induced double cascades, we expect a correlation between the energy of the event and the tau decay length. Events stemming from other flavors typically cluster at the thresholds, both in total deposited energy due to the falling spectrum and in double cascade length due to the very small mean reconstructed double cascade length for true single cascades. The total likelihood for the three topologies single cascade, track, and double cascade is $\mathcal{L} = \mathcal{L^{\rm SC}}  \mathcal{L^{\rm T}} \mathcal{L^{\rm DC}}$, where $\mathcal{L}^{\rm SC,T,DC}$ is the HESE binned-likelihood over the single cascade, track and double cascade events. 
While in \cite{HESE7}, the total likelihood is maximized assuming an equipartition of the flavors, here we fit the three flavors' individual contributions $f_{\alpha}$ to the overall astrophysical normalization. $f_{\alpha}$ is the fraction of $\nu_{\alpha}$ observed on Earth, with the constraint $f_e +f_{\mu} +f_{\tau} =1$. 
The result of the flavor composition measurement and the sensitivity are shown in Figure \ref{fig:flavor3}, the best-fit point is a composition of $\nuratio = 0.29:0.50:0.21$. Our fit flavor composition is consistent with previously published results by IceCube \cite{APJ15, Spencer, Marcel}, as well as with the expectation of $\sim$ 1 : 1 : 1 flavor composition on Earth coming from a pion decay production mechanism. It is also consistent with a zero astrophysical $\nutau$\, component.
The double cascade events' observables are shown in Table \ref{tab:DC}. ``Big Bird'' (Event\#1) has a large positive energy asymmetry, very close to the cut value of 0.3 where the background contribution from single cascades is significant. Figure \ref{fig:EA} shows the energy asymmetry for the best fit spectrum. It can be seen that ``Big Bird'' is in a background dominated region while in the case of ``Double Double'' (Event\#2), the observables are in a signal-dominated region. To firmly conclude how compatible each of the double cascades is with a background hypothesis, i.e. with not being due to a $\nu_{\tau}$-CC interaction, an a posteriori analysis was performed.

\begin{table}[htb]
\centering
\caption{Observables of the two Double Cascades}
\begin{tabular}{l | ll}
  & Event\#1 & Event\#2  \\
  & (Big Bird) & (Double Double) \\\hline
Energy of 1st cascade & 1.2 PeV & 9 TeV \\
Energy of 2nd cascade & 0.6 PeV & 80 TeV \\
Energy Asymmetry & 0.29 & -0.80 \\
Length & 16 m & 17 m \\ \hline
\end{tabular}
\caption{Observables of the two Double Cascades. ``Big Bird'' has a large positive energy asymmetry, very close to the cut at 0.3.}
\label{tab:DC}
\end{table}

\section{Detailed study of candidate events}
\begin{table}[htb]
\centering
\begin{tabular}{l | cc}
Variable & ``Big Bird'' & ``Double Double''  \\\hline
Primary Energy  & >65~TeV & >1.5~PeV \\
Visible Energy & 60 - 300~TeV & 1 - 3~PeV \\ 
Vertex, $r-r_{\mathrm{evt}}$ & \multicolumn{2}{c}{50~m } \\
Vertex, $z-z_{\mathrm{evt}}$ &  \multicolumn{2}{c}{25~m}    \\
Azimuth $\phi - \phi_{\mathrm{evt}}$ &  \multicolumn{2}{c}{$\sim 110\degree$ }  \\ 
Zenith $\theta - \theta_{\mathrm{evt}}$ &  \multicolumn{2}{c}{$\sim 35 \degree$}   \\ \hline
\end{tabular}
\caption{Restricted parameter space for resimulated events. The upper value of the primary energy depends on the interaction type, reflecting the spread of visible energy losses typical for that interaction. $r-r_{\mathrm{evt}}$ is the two-dimensional distance in the $x,y$-plane. For ``Big Bird''-like events, the zenith and azimuth regions were further restricted for $\numu$-CC interactions to maximize the number of events with reconstructed values close to the observed ones, as the direction resolution is very good for tracks.}
\label{tab:resim}     
\end{table}
The MC statistics available at the time of the unblinding were absolutely insufficient to incorporate the full information from the energy asymmetry observable. Further, as can be seen from Fig. \ref{fig:PID}, ``Big Bird'' lies in a region with very limited MC statistics, and ``Double Double'' lies in a region with both signal and background according to the MC templates. In order to incorporate the energy asymmetry observable, and to properly evaluate the signal and background contributions for the tau candidate events observed, a dedicated resimulation of each of the events was performed in a restricted parameter space, shown in Table \ref{tab:resim}. The mapping between true and reconstructed quantities is not straightforward. The interaction vertex in the resimulation was restricted to a cylinder with radius 50\,m and height 50\,m, the direction of the incoming neutrino spans $70 \degree$ in zenith and $220 \degree$ in azimuth, centered on the reconstructed interaction vertex and direction of the events. 
For the zenith and azimuth angles, the resolution depends on the event topology. While the azimuth was chosen to cover a wide range to account for possible contributions from azimuthal regions affected by the ice anisotropy and due to the limited azimuthal resolution for single cascades, the zenith region was restricted slightly more as the zenith resolution is better. With an angular resolution of $\sim 10 \degree$ for cascades, this seemed sufficient. However, the simulation did not include very down-going events, which could contribute up to $\sim3\%$ to the signal and $\sim 6 \%$ to the single cascade background classified as double cascades. In case of the primary energy, the mapping depends on the neutrino spectrum and the interaction type, and is only well-correlated to the reconstructed deposited energy for $\nue$-CC interactions, as only in this case the neutrino deposits its entire energy in the form of visible energy in the detector. All other interactions have some non-visible energy losses -- final state neutrinos, intrinsically darker hadronic cascades, muons leaving the detector -- such that it is not a priori known what primary energy range will significantly contribute to the region around the reconstructed values. The primary neutrino energy was restricted such as to cover all energies that can contribute 
to the observed reconstructed energies in reconstructed space, which had to be determined by trial and error. True quantities for the energy asymmetry and double cascade length are only defined for $\nutau$-CC interactions. Those variables were left unconstrained during resimulation.

\subsection{The expected fraction of $\nutau$-CC events}
The posterior probability for each event to have originated from a $\nutau$ CC interaction is:
\begin{equation}
P(\nutau | \vec \eta_{\rm{evt}}) = \frac{P_{\nutau}(\vec \eta_{\rm{evt}}) \; N^\mathrm{DC}_{\nutau}}{P_{\nutau}(\vec \eta_{\rm{evt}}) \; N^\mathrm{DC}_{\nutau} \; + \; P_{\cancel{\nutau}}(\vec \eta_{\rm{evt}}) \; N^\mathrm{DC}_{\cancel{\nutau}}} ,
\label{eq:tauness}
\end{equation}
where $N_{\nutau}$ and $N_{\cancel{\nutau}}$ are the total expected number of events in the double cascade bin stemming from $\nutau$-CC and non-$\nutau$-CC interactions and taken from the total simulation. $P_{\nutau}$ and $P_{\cancel{\nutau}}$ are the PDFs for the $\nutau$-CC and non-$\nutau$-CC components in the parameter space vector of each event, $\vec \eta_{\rm{evt}}$. We can approximate this expression as the ratio of the differential rates at $\vec \eta_{\rm{evt}}$,
\begin{equation}
P(\nutau | \vec \eta_{\rm{evt}}) \approx \; \frac{ \diff{N^\mathrm{DC}_{\nutau}} / \diff{\vec \eta_{\rm{evt}}}}{\diff{N^\mathrm{DC}_{\nutau}} / \diff{\vec \eta_{\rm{evt}}} + \diff{N^\mathrm{DC}_{{\cancel{\nutau}}}} / \diff{\vec \eta_{\rm{evt}}}} \; \equiv \; \tau.
\label{eq:tauness2}
\end{equation}
In the last step we have defined the tauness $\tau$, the fraction of events close to the reconstructed observables $\vec \eta_{\rm{evt}}$ of the data events, which are expected to be of the $\nutau$-CC interaction type. Note that the tauness is always evaluated under certain assumptions for the physics parameters $\vec \theta$, e.g. for $\vec \theta = \hat {\vec \theta}$ maximizing $\mathcal{L} = \mathcal{L^{\rm C}} \mathcal{L^{\rm T}} \mathcal{L^{\rm DC}}$. We adopt this assumption in the following, making the prior probability for a $\nutau$ interaction to be the $\nutau$ content in the double cascade topology subsample according to the best-fit with a 1:1:1 composition \cite{HESE7}. Substituting $\diff{N^\mathrm{DC}_{\nutau}} / \diff{\vec \eta_{\rm{evt}}} = N_{\nutau} P_{\nutau}(\vec \eta_{\rm{evt}})$, where $N_{\nutau} P_{\nutau}(\vec \eta_{\rm{evt}})$ is the differential expected number of $\nutau$-CC events at the point $\vec \eta_{\rm{evt}}$, and similarly for non-$\nutau$-CC interactions ($\cancel{\nutau}$), we can write
\begin{equation}
  \tau  = \frac{N_{{\nutau}} P_{{\nutau}}(\vec \eta_{\rm{evt}})}{N_{\nutau} P_{\nutau}(\vec \eta_{\rm{evt}}) + N_{\cancel{\nutau}} P_{\cancel{\nutau}}(\vec \eta_{\rm{evt}})}.
\end{equation}
The differential expected number of events at the point $\vec \eta_{\rm{evt}}$, $N_{\nutau} P_{\nutau}(\vec \eta_{\rm{evt}})$ and $N_{\cancel{\nutau}} P_{\cancel{\nutau}}(\vec \eta_{\rm{evt}})$ can be computed from the resimulation sets using a multidimensional KDE with a gaussian kernel and the Rodeo algorithm \cite{rodeo}. The dimensionality necessarily needs to be extended to more than the 2 dimensions used in the original double cascade classification before, as the resimulation was carried out in the restricted parameter space. The additional dimensions are 3 dimensions for the vertex \textit{x, y, z} and 2 dimensions for the direction $\theta, \phi$. With the total deposited energy we have 6 resulting dimensions. We further define a region of interest in the parameters not restricted during resimulation, double cascade length and energy asymmetry. 

\subsection{Results from the a-posteriori evaluation}
The dedicated resimulation for the a posteriori evaluation yields a total of $\sim 13 \cdot 10^6$ ``Double-Double''-like events passing the HESE selection and $\sim 1 \cdot 10^6$ ``Big-Bird''-like events passing the HESE selection. We compute the tauness for each of the events to stem from a $\nutau$-CC interaction for the best-fit spectrum and $1:1:1$ flavor composition as given in \cite{HESE7}. The tauness for ``Big Bird'' is $\tau_{\mathrm{best\,fit}}^{\mathrm{BB}} \approx 75\%$, the tauness for ``Double Double'' is $\tau_{\mathrm{best\,fit}}^{\mathrm{DD}} \gtrsim 97\%$. That means, that we expect $\sim 75\, \, (\gtrsim97)$ out of 100 observed events with Big Bird (Double Double) observables to be stemming from $\nutau$-CC interactions. For Double Double, the statistics of the generated MC are not sufficient to evaluate the tauness to a higher precision. Note that the uncertainties in the fit spectrum and flavor composition have not yet been taken into account. For both events, the major background component misclassified as double cascades are true single cascades, predominantly from $\nue$ interactions. In the evaluated region of interest, the single cascade background is strongly suppressed, but still dominant for ``Big Bird''  for which single cascades are making up $\gtrsim 4/5$ of the background. Note also, that while for ``Big Bird'' the single cascade hypothesis fit yields a satisfying result, ``Double Double'' cannot be described well with a single cascade hypothesis fit. 

\section{Summary and Discussion}
The latest update to the HESE sample and analyses with 7.5 years of livetime includes a dedicated algorithm to determine the topology of the observed events and can distinguish between single cascades, double cascades, and tracks. Double cascades are created by $\nutau$-CC interactions, but due to the fast decay of the produced tau lepton, they are very difficult to identify. We have for the first time classified two of the IceCube data events as double cascades, thus finding promising candidates for astrophysical tau neutrinos. An a-posteriori analysis was performed to better assess the probability of each of the candidates to be stemming from a $\nutau$-CC interaction, based on newly produced MC events covering a similar parameter space as the observed data. For the single power-law spectrum fit to the HESE sample, we find that ``Double Double'' has a tauness of $\gtrsim 0.97$ and ``Big Bird'' has a tauness of $\approx 0.75$. This means that in a very long-running experiment we would expect at most one event for every 30 events observed with reconstructed values similar to ``Double Double'' to be due to a non-$\nutau$-CC interaction, while a background scenario would account for one out of every four events seen with reconstructed values similar to ``Big Bird''. 
Even with the resimulated MC events, the MC statistics are still the limiting factor for Double Double. 
In the meantime, two complementary tau searches were performed. Looking for two resolved pulses on single sensors, they use different information than the double cascade method discussed here. Both analyses \cite{DP1,DP2} also find ``Double Double''. 

\bibliographystyle{ICRC}
\bibliography{proceedings}

\end{document}